\documentclass[aps,preprint,showpacs,showkeys]{revtex4}%
\usepackage{amsfonts}
\usepackage{amsmath}
\usepackage{amssymb}
\usepackage{graphicx}%
\begin{document}
\title{Genetic algorithm for the pair distribution function of the electron gas}
\author{Fernando Vericat}
\altaffiliation{Also at Instituto de F\'{\i}sica de L\'{\i}quidos y Sistemas Biol\'{o}gicos
(IFLYSIB)-CONICET-CCT La Plata, Argentina.}

\altaffiliation{E-mail: vericat@iflysib.unlp.edu.ar}

\affiliation{Grupo de Aplicaciones Matem\'{a}ticas y Estad\'{\i}sticas de la Facultad de
Ingenier\'{\i}a (GAMEFI). Universidad Nacional de La Plata, Argentina}
\author{C\'{e}sar O. Stoico}
\affiliation{Area F\'{\i}sica, Facultad de Ciencias Bioqu\'{\i}micas y Farmac\'{e}uticas,
Universidad Nacional de Rosario, Argentina.}
\author{C. Manuel Carlevaro}
\affiliation{Instituto de F\'{\i}sica de L\'{\i}quidos y Sistemas Biol\'{o}gicos
(IFLYSIB)-CONICET- CCT La Plata, Argentina.}
\author{Danilo G. Renzi}
\affiliation{Facultad de Ciencias Veterinarias, Universidad Nacional de Rosario, Casilda, Argentina}

\begin{abstract}
The pair distribution function of the electron gas is calculated using a
parameterized generalization of quantum hypernetted chain approximation with
the parameters being obtained by optimizing the system energy with a genetic
algorithm. The functions so obtained are compared with Monte Carlo simulations
performed by other authors in its variational and diffusion versions showing a
very good agreement especially with the diffusion Monte Carlo results.\bigskip

\end{abstract}

\maketitle

\section{Introduction}

The Sommerfeld electron gas model\cite{Hoddeson1} has proved to be very useful
in describing many of the electronic properties of metallic solids. It
represents the conduction electrons as a zero temperature ensemble of point
charged fermions moving against a continuous neutralizing background that
plays the role of the ionic lattice. In the simplest version of the model,
fermions are considered spinless and the background is just characterized by a
dielectric constant. If we have $N$ fermions of mass $m$ each one carrying a
charge $e$, then the system Hamiltonian reads%

\begin{equation}
H=%
\genfrac{(}{)}{}{}{\hbar^{2}}{2m}%
{\displaystyle\sum\limits_{i=1}^{N}}
\nabla_{i}^{2}+%
{\displaystyle\sum\limits_{i<j}}
v\left(  r_{ij}\right)  . \tag{1}\label{1}%
\end{equation}
Here $v\left(  r_{ij}\right)  $ is the pair potential given by%

\begin{equation}
v\left(  r_{ij}\right)  =\frac{e^{2}}{\varepsilon r_{ij}}, \tag{2}\label{2}%
\end{equation}
where $r_{ij}=\left\vert \mathbf{r}_{i}-\mathbf{r}_{j}\right\vert $ \ with
$\mathbf{r}_{i\text{ }}$ the position of the $i$th-particle and $\varepsilon$
denoting the dielectric constant.

The equilibrium behavior of this system has been widely studied through
quantal Monte Carlo simulations\cite{Ceperley1},\cite{Ceperley2} and also from
a variety of many-body theories\cite{March1}-\cite{Kraeft1}. Many of them
center on the pair distribution functions (PDF)\cite{Singwi1}. If the we
denote $\psi\left(  \mathbf{r}_{1\text{ }},\mathbf{r}_{2\text{ }}%
,\cdots,\mathbf{r}_{N\text{ }}\right)  $ the $N$-body wave function then the
pair distribution function is given%

\begin{align}
\rho\left(  \mathbf{r}_{1\text{ }},\mathbf{r}_{2\text{ }}\right)   &
=\rho\left(  \mathbf{r}_{1\text{ }}\right)  \rho\left(  \mathbf{r}_{2\text{ }%
}\right)  g\left(  \mathbf{r}_{1\text{ }},\mathbf{r}_{2\text{ }}\right)
\nonumber\\
&  =N\left(  N-1\right)  \frac{%
{\displaystyle\idotsint}
d\mathbf{r}_{3\text{ }}d\mathbf{r}_{4\text{ }}\cdots d\mathbf{r}_{N\text{ }%
}\left\vert \psi\left(  \mathbf{r}_{1\text{ }},\mathbf{r}_{2\text{ }}%
,\cdots,\mathbf{r}_{N\text{ }}\right)  \right\vert ^{2}}{%
{\displaystyle\idotsint}
d\mathbf{r}_{1\text{ }}d\mathbf{r}_{2\text{ }}\cdots d\mathbf{r}_{N\text{ }%
}\left\vert \psi\left(  \mathbf{r}_{1\text{ }},\mathbf{r}_{2\text{ }}%
,\cdots,\mathbf{r}_{N\text{ }}\right)  \right\vert ^{2}} \tag{3}\label{3}%
\end{align}
where the integrations are over the whole volume and%

\begin{equation}
\rho\left(  \mathbf{r}_{1}\right)  =N\frac{%
{\displaystyle\idotsint}
d\mathbf{r}_{2\text{ }}d\mathbf{r}_{3\text{ }}\cdots d\mathbf{r}_{N\text{ }%
}\left\vert \psi\left(  \mathbf{r}_{1\text{ }},\mathbf{r}_{2\text{ }}%
,\cdots,\mathbf{r}_{N\text{ }}\right)  \right\vert ^{2}}{%
{\displaystyle\idotsint}
d\mathbf{r}_{1\text{ }}d\mathbf{r}_{2\text{ }}\cdots d\mathbf{r}_{N\text{ }%
}\left\vert \psi\left(  \mathbf{r}_{1\text{ }},\mathbf{r}_{2\text{ }}%
,\cdots,\mathbf{r}_{N\text{ }}\right)  \right\vert ^{2}} \tag{4}\label{4}%
\end{equation}
is the one point distribution function. The function $g\left(  \mathbf{r}%
_{1\text{ }},\mathbf{r}_{2\text{ }}\right)  $ is the pair correlation
function\ (PCF). For homogenous systems (as will be \ considered here)
$\rho\left(  \mathbf{r}_{i\text{ }}\right)  $ gives the electrons number
density $\rho\left(  \mathbf{r}_{i\text{ }}\right)  =\rho=N/V$ ($V=$ system
volume) and the PDF and PCF depend only on the particles separation:
$\rho\left(  r_{12}\right)  =\rho^{2}g\left(  r_{12}\right)  $ .

For the homogeneous electron gas in three and lower dimensions, the PDF as
well its Fourier transform, the static structure factor $S\left(  k\right)  $,
have been studied by several authors using diverse analytical
techniques\cite{Mahan1} and also simulations methods. In 3D, we mention random
phase approximation (RPA) calculations\cite{Pines1}, diagrammatic ladder
approximations\cite{Yasuhara1}, the local-field based method of Singwi, Tosi,
Land and Sj\"{o}lander (STLS)\cite{Singwi2}, calculations with quantum
hypernetted chain equations (QHNC)\cite{Lantto1} and also techniques that use
parameterized PDF, the parameters being determined from known independent
theoretical or simulation results\cite{Rajagopal1}-\cite{GoriGiorgi1}. From
the side of Monte Carlo quantum simulations in both -the variational and
diffusion- versions, the work by Ortiz and Ballone\cite{Ortiz1} extends in
several ways previous results of Ceperley and Alder\cite{Ceperley1}%
,\cite{Ceperley2}.

Most of these approaches lean on the variational principle according to which
the energy of the ground state $E_{0}$ is a lower bound for the Hamiltonian
mean value as calculated using any trial wave function $\psi_{T}\left(
\mathbf{r}_{1\text{ }},\mathbf{r}_{2\text{ }},\cdots,\mathbf{r}_{N\text{ }%
}\right)  $:%

\begin{equation}
\frac{\left\langle \psi_{T}\right\vert H\left\vert \psi_{T}\right\rangle
}{\left\langle \psi_{T}\right\vert \left.  \psi_{T}\right\rangle }\geq E_{0}.
\tag{5}\label{5}%
\end{equation}
As trial function a factorized form%

\begin{equation}
\psi_{T}\left(  \mathbf{r}_{1\text{ }},\mathbf{r}_{2\text{ }},\cdots
,\mathbf{r}_{N\text{ }}\right)  =F\left(  \mathbf{r}_{1\text{ }}%
,\mathbf{r}_{2\text{ }},\cdots,\mathbf{r}_{N\text{ }}\right)  \psi_{0}\left(
\mathbf{r}_{1\text{ }},\mathbf{r}_{2\text{ }},\cdots,\mathbf{r}_{N\text{ }%
}\right)  \tag{6}\label{6}%
\end{equation}
is frequently used. Here $\psi_{0}\left(  \mathbf{r}_{1\text{ }}%
,\mathbf{r}_{2\text{ }},\cdots,\mathbf{r}_{N\text{ }}\right)  $ denotes the
system wave function when the interactions are turned off ($v\left(
r_{ij}\right)  \equiv0$). It is an antisymmetric function under particles
permutations. We can write%

\begin{equation}
\psi_{0}\left(  \mathbf{r}_{1\text{ }},\mathbf{r}_{2\text{ }},\cdots
,\mathbf{r}_{N\text{ }}\right)  =%
{\displaystyle\sum\limits_{P}}
\left(  -1\right)  ^{P}P\left\{  \phi_{1}\left(  \mathbf{r}_{1\text{ }%
}\right)  ,\phi_{2}\left(  \mathbf{r}_{2\text{ }}\right)  ,\cdots,\phi
_{N}\left(  \mathbf{r}_{N\text{ }}\right)  \right\}  =\det\left[  \phi
_{\alpha_{i}}\left(  \mathbf{r}_{j\text{ }}\right)  \right]  \tag{7}\label{7}%
\end{equation}
where $P$ is the permutation operator that interchanges the particle
positions, $\phi_{\alpha_{i}}\left(  \mathbf{r}_{j\text{ }}\right)  $ is the
wave function of an isolated particle and\ $\det\left[  \phi_{\alpha_{i}%
}\left(  \mathbf{r}_{j\text{ }}\right)  \right]  $ means the Slater
determinant. The symmetric factor $F\left(  \mathbf{r}_{1\text{ }}%
,\mathbf{r}_{2\text{ }},\cdots,\mathbf{r}_{N\text{ }}\right)  $ accounts for
the correlations among the particles when the interactions are turned on. The
$N$-body correlation factor can be factorized according to
Jastrow\cite{Jastrow1}:
\begin{equation}
F\left(  \mathbf{r}_{1\text{ }},\mathbf{r}_{2\text{ }},\cdots,\mathbf{r}%
_{N\text{ }}\right)  =%
{\displaystyle\prod\limits_{i<j}}
f\left(  \mathbf{r}_{i\text{ }},\mathbf{r}_{j\text{ }}\right)  .
\tag{8}\label{8}%
\end{equation}

In this work we consider the evaluation of the PDF for the homogeneous 3D
electron gas starting from a parameterized trial\ wave function of the form
given by Eqs. (\ref{6}-\ref{8}) with the parameters obtained by optimizing the
system energy by means of a genetic algorithm.

Genetic algorithms\cite{Goldberg1}-\cite{Mitchell1} form part, together with
evolutionary programming\cite{Fogel1},\cite{Back1}, game-playing
strategies\cite{Davis1}, genetic programming\cite{Koza1} and other related
techniques,\ of\ a relatively new class of optimization algorithms which are
based on the Darwinian evolution principle\cite{Holland1}. In particular,
genetic algorithms tackle even complex problems with surprising efficiency and
robustness. In Physics they have been used in calculations that involve from
simple quantum systems\cite{Grogorenko1} to astrophysical
systems\cite{Charbonneau1}, running through lattice models for spin
glasses\cite{Prugel1}, molecules\cite{Moret1} and clusters\cite{Zeiri1}. More
recently\cite{Stoico1} we have developed a genetic algorithm for the PDF of
the one-dimensional electron gas in what, at our knowledge, is the first
application of this kind of algorithm to describe many-body systems in the
thermodynamic limit..

In general, a genetic algorithm is based on three main statements:

a) It is a process that works at the chromosomic level. Each individual is
codified as a set of chromosomes.

b) The process follows the Darwinian theory of evolution, say, the survival
and reproduction of those individuals that best adapt in a changing environment.

c) The evolutionary process takes place at the reproduction stage. It is in
this stage when mutation and crossover occurs. As a result, the progeny
chromosomes can differ from their parents ones.

Starting from a guess initial population, a genetic algorithm basically
generates consecutive generations (offprints). These are formed by a set of
chromosomes, or character (genes) chains, which represent possible solutions
to the problem under consideration. At each algorithm step, a fitness function
is applied to the whole set of chromosomes of the corresponding generation in
order to check the goodness of the codified solution. Then, according to their
fitting capacity, couples of chromosomes, to which the crossover operator will
be applied, are chosen. Also, at each step, a mutation operator is applied to
a number of randomly chosen chromosomes.

The two most commonly used methods to randomly select the chromosomes are:

i) \textit{The roulette wheel algorithm}. It consists in building a roulette,
so that to each chromosome corresponds a circular sector proportional to its fitness.

ii) \textit{The tournament method}. After shuffling the population, their
chromosomes are made to compete among them in groups of a given size
(generally in pairs). The winners will be those chromosomes with highest
fitness. If we consider a binary tournament, say the competition is between
pairs, the population must be shuffled twice. This technique guarantees copies
of the best individual among the parents of the next generation.

After this selection, we proceed with the sexual reproduction or crossing of
the chosen individuals. In this stage, the survivors exchange chromosomic
material and the resulting chromosomes will codify the individuals of the next
generation. The forms of sexual reproduction most commonly used are:

i) With one crossing point. This point is randomly chosen on the chain length,
and all the chain portion between the crossing point and the chain end is exchanged.

ii)\ With two crossing points. The portion to be exchanged is in between two
randomly chosen points.

For the algorithm implementation, the crossover normally has an assigned
percentage that determines the frequency of its occurrence. This means that
not all of the chromosomes will exchange material but some of them will pass
intact to the next generation. As a matter of fact, there is a technique,
named elitism, in which the fittest individual along several generations does
not cross with any of the other ones and keeps intact until an individual
fitter than itself appears.

Besides the selection and crossover, there is another operation, mutation,
that produces a change in one of the\ characters or genes of a randomly chosen
chromosome. This operation allows to introduce new chromosomic material into
the population. As for the crossover, the mutation is handled as a percentage
that determines its occurrence frequency. This percentage is, generally, not
greater than 5\%, quite below the crossover percentage.

Once the selected chromosomes have been crossed and muted, we need some
substitution method. Namely,\ \ we must choose, among those individuals, which
ones will be substituted for the new progeny. Two main substitution ways are
usually considered. In one of them, all modified parents are substituted for
the generated new individuals. In this way an individual does never coexist
with its parents. In the other one, only the worse fitted individuals of the
whole population are substituted, thus allowing the coexistence among parents
and progeny.

Since the answer to our problem is almost always unknown, we must establish
some criterion to stop the algorithm. We can mention two such criteria: i) the
algorithm is run along a maximum number of generations; ii) the algorithm is
ended when the population stabilization has been reached, i.e. when all, or
most of, the individuals have the same fitness.

In Section III we will apply these ideas to determine the parameters appearing
in the expression for the 3D electron gas PDF that we propose in Section II.

\section{Parameterized PDF}

We assume that the trial wave function for the system of $N$-spinless
electrons with Hamiltonian given by Eqs.(\ref{1},\ref{2}) has the form of
Eq.(\ref{6}) where we use for the ideal part $\psi_{0}\left(  \mathbf{r}%
_{1\text{ }},\mathbf{r}_{2\text{ }},\cdots,\mathbf{r}_{N\text{ }}\right)  $ a
parameterized generalization of an expression given by Lado\cite{Lado1} and
for the Jastrow correlation factor also a parameterized expression containing
a random phase approximation (RPA) pseudopotential\cite{Ceperley1}%
,\cite{Gaskell1}. Specifically we propose%

\begin{equation}
\psi_{T}\left(  \mathbf{r}_{1\text{ }},\mathbf{r}_{2\text{ }},\cdots
,\mathbf{r}_{N\text{ }}\right)  =\exp\left\{  \frac{1}{2}%
{\displaystyle\sum\limits_{i<j}}
\left[  \alpha w_{0}\left(  r_{ij}\right)  -\beta u_{RPA}\left(
r_{ij}\right)  \right]  \right\}  \tag{9}\label{9}%
\end{equation}
where $\alpha$ and $\beta$ are the parameters to adjust.

In Eq.(\ref{9}) the ideal gas effective potential $w_{0}\left(  r\right)  $ is defined%

\begin{equation}
w_{0}\left(  r\right)  =\ln g_{0}\left(  r\right)  -\frac{1}{\left(
2\pi\right)  ^{3}\rho}%
{\displaystyle\int}
d\mathbf{k}e^{-i\mathbf{k.r}}\frac{\left[  S_{0}\left(  k\right)  -1\right]
^{2}}{S_{0}\left(  k\right)  }. \tag{10}\label{10}%
\end{equation}
Here $g_{0}\left(  r\right)  $ and $S_{0}\left(  k\right)  $ are the ideal PCF
and structure factor, respectively, whose expressions are\cite{Placzek1}%
,\cite{London1}:%

\begin{equation}
g_{0}\left(  r\right)  =1-\frac{9}{2}\left[  \frac{\sin\left(  k_{F}r\right)
-k_{F}r\cos\left(  k_{F}r\right)  }{\left(  k_{F}r\right)  ^{3}}\right]  ^{2}
\tag{11}\label{11}%
\end{equation}
and%

\begin{equation}
S_{0}\left(  k\right)  =%
\genfrac{\{}{.}{0pt}{}{\text{ \ \ \ \ \ \ \ \ }1\text{
\ \ \ \ \ \ \ \ \ \ \ \ \ \ \ \ \ \ \ \ \ }k>2k_{F}}{\frac{3}{4}\frac{k}%
{k_{F}}-\frac{1}{16}\left(  \frac{k}{k_{F}}\right)  ^{3}\text{ \ \ \ \ \ \ }%
k<2k_{F}\text{\ \ }}
\tag{12}\label{12}%
\end{equation}
with $k_{F}$ the Fermi momentum $k_{F}=\left(  3\pi^{2}\rho^{\ast}\right)
^{\frac{1}{3}}$ where $\rho^{\ast}=\rho a_{0}$ ($a_{0}$ is the Bohr radius).

The RPA pseudopotential, on the other hand, reads%

\begin{equation}
2\rho^{\ast}u_{RPA}\left(  k\right)  =-\frac{1}{S_{0}\left(  k\right)
}+\left[  \frac{1}{S_{0}\left(  k\right)  ^{2}}+\frac{4m\rho^{\ast}%
\widetilde{v}\left(  k\right)  }{\hbar^{2}k^{2}}\right]  ^{1/2} \tag{13}%
\label{13}%
\end{equation}
where $\widetilde{v}\left(  k\right)  $ is the Fourier transform of the
interparticle potential $v\left(  r\right)  $.

It is worth mentioning that in the genetic algorithm me have develop for the
1D electron gas in Ref.\cite{Stoico1}, instead of using the RPA
pseudopotential in Eq.(\ref{9}) we assume that $u\left(  r\right)  $ is an
unknown function and the algorithm is designed to completely obtain it.
Here,\textit{ }a parameterized form for the pseudopotential is proposed
\textit{a priori} and the algorithm looks for the optimal parameters.

In principle, to calculate PDF from Eq.(\ref{3}) we have to integrate the
square of the wave function over $3N$ coordinates. To avoid this formidable
task use is done of a modified form of the hypernetted chain approximation
(HNC) for which the PDF is written as a single integral equation involving
just a pair of particles. We write, ignoring all the elemental
diagrams\cite{Stoico2},%

\begin{equation}
g\left(  r\right)  =\exp\left[  \alpha w_{0}\left(  r\right)  -\beta
u_{RPA}\left(  r\right)  +N\left(  r\right)  \right]  \tag{14}\label{14}%
\end{equation}

\begin{equation}
N\left(  r\right)  =\rho%
{\displaystyle\int}
\left[  g\left(  r^{\prime}\right)  -1-N\left(  r^{\prime}\right)  \right]
\left[  g\left(  \left\vert \mathbf{r-r}^{\prime}\right\vert \right)
-1\right]  d\mathbf{r} \tag{15}\label{15}%
\end{equation}
where $N\left(  r\right)  $ denotes the sum of nodal diagrams. Eq. (\ref{14})
without the term with $\alpha$ in the exponential has the form of the PCF for
a system of bosons (see v.g. Eq.(40) in Ref. \cite{Stoico2} where the nodal
diagrams are denoted $D$ and the elementals diagrams $E$ must be\ taken zero).
The term with $\alpha$ adds the ideal part that contains the proper symmetry
for fermions.

Finally, in the variational approach which is implicit in Eq.(\ref{5}), we
need an expression for the Hamiltonian mean value. Making use of
Jackson-Feenberg identity\cite{Feenberg1} we obtain%

\begin{equation}
\frac{E}{N}=\frac{\rho}{2}%
{\displaystyle\int}
d\mathbf{r}g\left(  r\right)  \left[  -\frac{\hbar^{2}}{4m}\nabla^{2}\ln
f^{2}\left(  r\right)  +v\left(  r\right)  \right]  . \tag{16}\label{16}%
\end{equation}

\section{Parameters optimization}

Our problem is to solve for $g\left(  r\right)  $ the integral equation given
by Eqs.(\ref{14} and \ref{15}) with the parameters $\alpha$ and $\beta$
determined by demanding that the energy functional $E=E\left\{  g\left(
r\right)  \right\}  $\ given by Eq.(\ref{16}) be minimum. To this end we use a
genetic algorithm.

We proceed by first generating the initial population, which is formed by
$N_{p}$ random replicas of the two numbers string (that represents one
population individual) $\ \gamma^{\left(  \alpha\right)  },\gamma^{\left(
\beta\right)  }$ where $\gamma^{\left(  \alpha\right)  }\in\left[  0,1\right]
$ is a random real number (rounded to an established number $n$ of decimals)
which is assigned to the parameter $\alpha$. Given a string $\gamma^{\left(
\alpha\right)  },\gamma^{\left(  \beta\right)  }$, the \textit{encoding}
consists in replacing the sequence of real numbers by a single natural number
obtained by putting their decimals parts one next to the other. Thus if
$\gamma^{\left(  \alpha\right)  }=0.\gamma_{1}^{\left(  \alpha\right)  }$
$\gamma_{2}^{\left(  \alpha\right)  }...\gamma_{n}^{\left(  \alpha\right)  }$
and $\gamma^{\left(  \beta\right)  }=0.\gamma_{1}^{\left(  \beta\right)  }$
$\gamma_{2}^{\left(  \beta\right)  }...\gamma_{n}^{\left(  \beta\right)  }$
then we have the chain :
\[
\gamma_{1}^{\left(  \alpha\right)  }\gamma_{2}^{\left(  \alpha\right)
}...\gamma_{n}^{\left(  \alpha\right)  }\gamma_{1}^{\left(  \beta\right)
}\gamma_{2}^{\left(  \beta\right)  }...\gamma_{n}^{\left(  \beta\right)  }%
\]
and the population is the set%

\[
\left\{  \left(  \gamma_{1}^{\left(  \alpha\right)  }\gamma_{2}^{\left(
\alpha\right)  }...\gamma_{n}^{\left(  \alpha\right)  }\gamma_{1}^{\left(
\beta\right)  }\gamma_{2}^{\left(  \beta\right)  }...\gamma_{n}^{\left(
\beta\right)  }\right)  _{r}\text{ \ \ \ \ }r=1,2,\cdots,N_{p}\right\}
\]

In genetic terms, the encoding produces the chromosomic structure of the
individuals (replica string). The inverse process is called \textit{decoding}
and returns the parameters $\alpha$ and $\beta$ corresponding to each
individual. In the decoding we allow the returned parameters be multiplied by
a constant factor $\eta>1$ in order that the parameters can take values in an
interval wider than $\left[  0,1\right]  $. We define the fitness of the $r$th
individual as $f_{r}=e^{-E_{r}}$ where $E_{r}$ is the energy calculated using
Eq.(\ref{16}) when $g\left(  r\right)  $ is calculated from Eqs.(\ref{14} and
\ref{15}) for the parameters $\alpha$ and $\beta$ obtained by decoding the
chromosome structure of the $r$th individual of the population. A solution is
reached when\ $f_{r}\approx1$ for some individual $r$ in some of the
successive populations obtained in the evolution process.

The calculation proceeds by dividing the population of $N_{p}$ replicas into
$N_{p}/2$ couples. The couples are randomly chosen by using the roulette wheel
algorithm\cite{Goldberg1}. This is done by defining the sums $F=%
{\textstyle\sum\nolimits_{r=1}^{N_{p}}}
f_{r}$ and $S_{\delta}=%
{\textstyle\sum\nolimits_{r=1}^{\delta}}
$ $f_{r}$ \ ($\delta=1,2,...,N_{p}$). Then, a random number $\kappa\in\left[
0,F\right]  $ is generated and the unique index $\delta$ such that
$S_{\delta-1}\leq\kappa\leq S_{\delta\text{ }}$is picked up.

Once the first generation of replicas (parents) has been generated and divided
into couples, the second generation (offspring) can be generated by applying
the crossover operator between the members of each one. At times , some of the
members of the new replicas generation can be changed by applying the mutation operator.

Given a couple of replicas, the crossover operator is defined by generating a
new random number $c\in\left[  0,1\right]  $ which is compared with a
pre-established crossover probability $p\in\left[  0,1\right]  $. If $c\leq
p$, the crossover operator acts by interchanging all the digits from the $s$th
position to the end of the replica between the members of the couple.\ Here
$s$ is a random integer such that $1\leq s\leq2n$. for example if the couple is

$153280472...337$

$768325399...069$

and $s=4,$ the new offspring couple will be

$153225399...069$

$768380472...337.$

To apply the mutation operator we first randomly select those offsprings that
will mutate. Then, for each of these offsprings, a gene (a digit) randomly
chosen is changed by a random integer number $\ell\in\left[  0,9\right]  .$The
algorithm is stopped when a solution is reached for the parameters $\alpha$
and $\beta$.

\section{Results}

\begin{table}
 \caption{Parameters $\alpha$ and $\beta$}
 \label{tbl:1}
\begin{tabular}
[c]{|c|c|c|}\hline
$r_{s}$ & $\ \alpha$ & $\ \beta$\\\hline
1 & 2.2227 & 1.6456\\\hline
2 & 7.8707 & 14.865\\\hline
3 & 9.6789 & 30.526\\\hline
6 & 14.710 & 43.030\\\hline
10 & 44.974 & 18.464\\\hline
50 & -6834.17 & 128.71\\\hline
\end{tabular}
\end{table}

As it is easily seen, the electron gas is completely determined by giving just
its density $\rho$ or, as it is custom in many-body theory, the Wigner-Seitz
radius $r_{s}$ defined $r_{s}=\left[  3/4\pi\rho\right]  ^{1/3}$. Here we have
applied the procedure described above, to an electron gas at metallic
densities: $1\leq r_{s}\leq10$ and also at $r_{s}=50$. We use in our
calculations $N_{p}=150$ and $n=5$. The factor $\eta$, in turn, is moved in
each case to give reasonable values for the parameters $\alpha$ and $\beta$.
In Table \ref{tbl:1} we show the parameters $\alpha$ and $\beta$ obtained for the
diverse values of $r_{s\text{ }}$considered. Note the change in the parameters
tendency at $r_{s}=50$. By the way, we were unable to reach convergence for
values of $r_{s}$ greater than this one.

Figs. 1 to 3 show the corresponding $g\left(  r\right)  ^{\prime}$s. In Fig. 1
we put all together the curves we have obtained . We observe the
characteristic features of the electron gas: when the density increases
($r_{s}$ decreases) the behavior tends to that of an ideal paramagnetic gas of
fermions with contact value $1/2$ and rapidly going to the asymptotic value
$1$. When the density decreases the correlation functions become more
structured showing, in particular, a more pronounced Coulomb hole near contact.

\begin{figure}[t]
 \centering
 \includegraphics[scale=0.5,keepaspectratio=true]{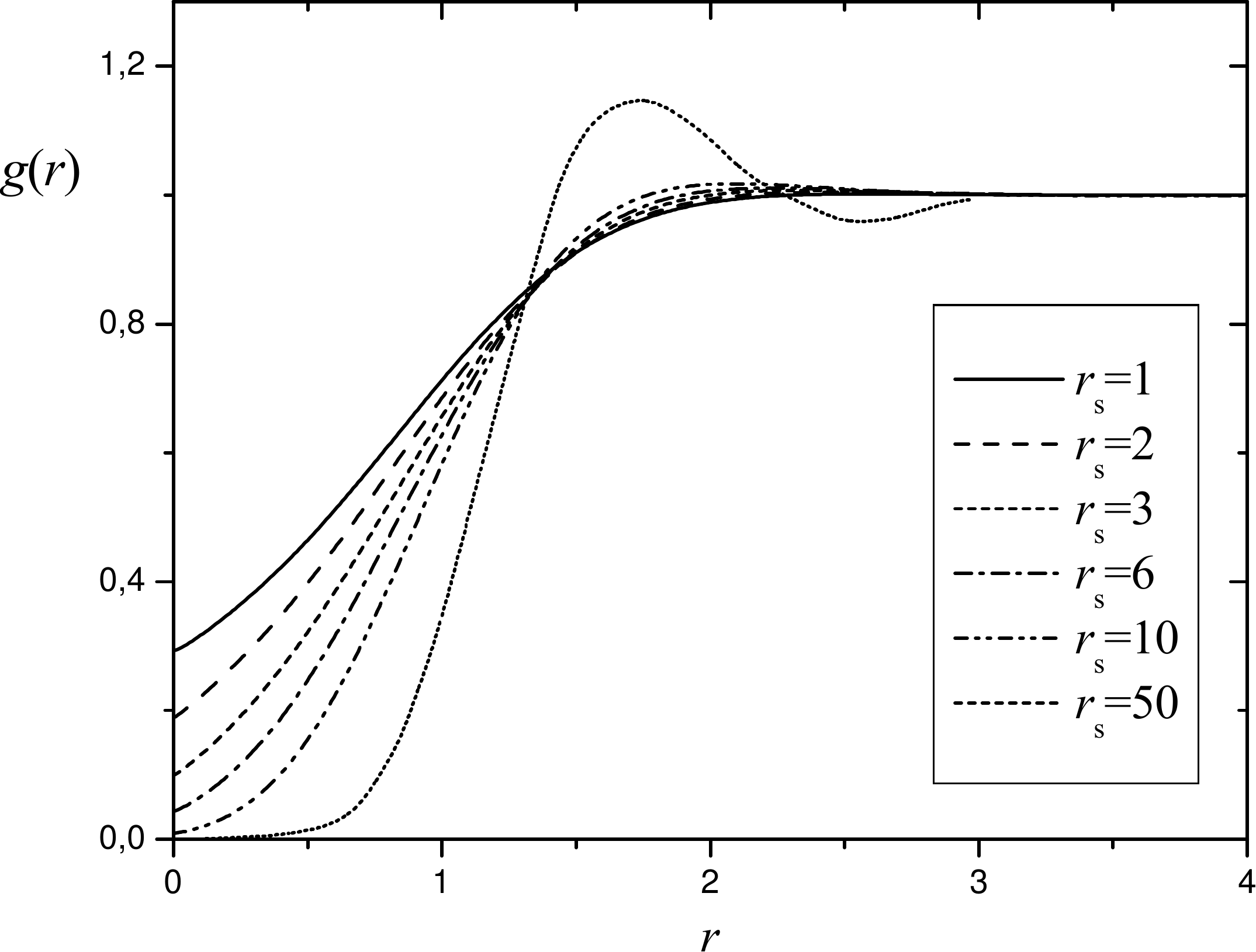}
 \caption{The functions $g\left(  r\right)  $ for the homogenous
electron gas at \textit{r}$_{s}$=1,2,3,6,10 and 50 obtained in this work.}
 \label{fig:1}
\end{figure}

In Figs. 2 and 3 comparison is done of our results with those obtained from
variational as well as diffusion Monte Carlo simulations performed by other
authors\cite{Ortiz1},\cite{Lantto1},\cite{Ceperley3}. A first remark is the
existing differences between variational and diffusion results particularly
for low values of $r_{s}$. Also must be noticed the good agreement of the
results of this work with those obtained from diffusion Monte Carlo calculations.

\begin{figure}[htb!]
 \centering
 \includegraphics[scale=0.65,keepaspectratio=true]{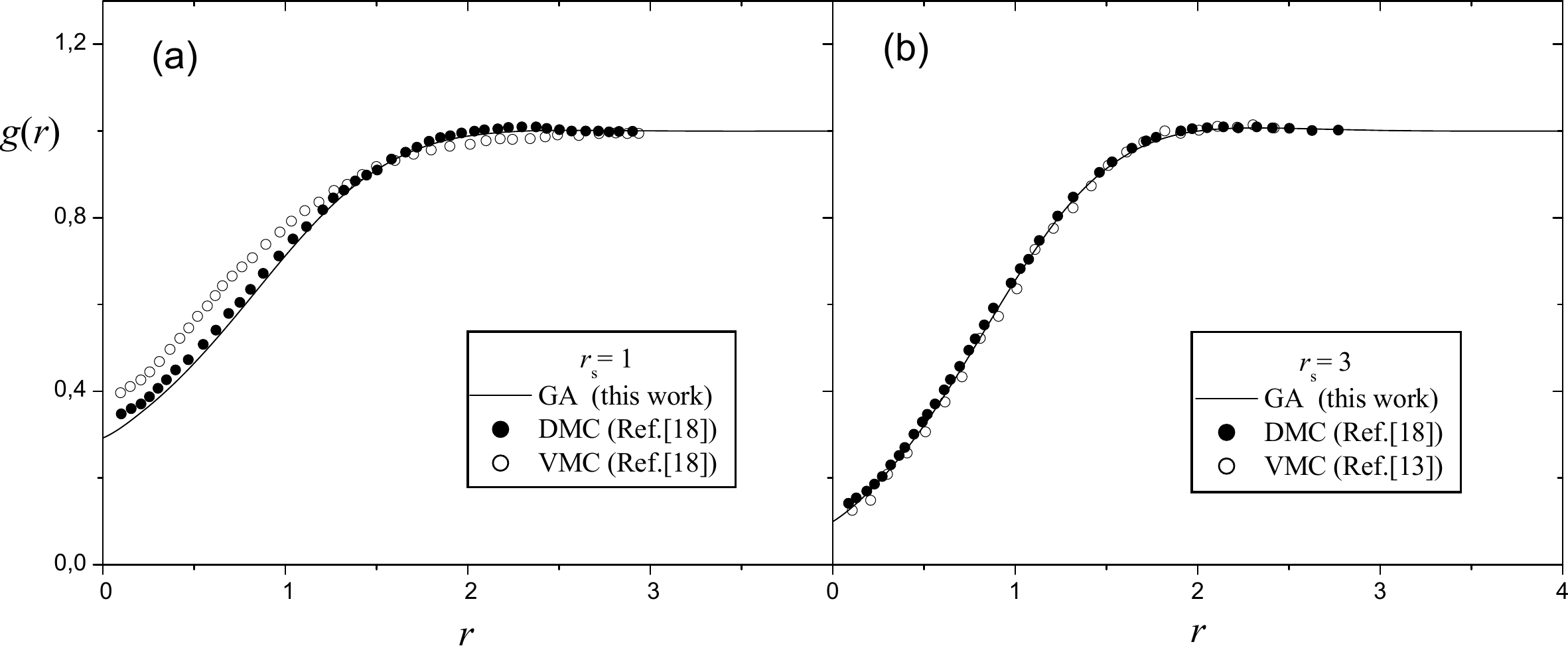}
 \caption{The functions $g(r)$ for the homogenous electron obtained
in this work compared with variational and diffusion Monte Carlo results. (a):
\textit{r}$_{s}$=1; (b): \textit{r}$_{s}$=3.}
 \label{fig:2-3}
\end{figure}

It is worth mentioning that the time to obtain one of our curves by running
the complete algorithm with a Pentium IV is tipically of the order of 50 hours
for the metallic densities.

\begin{figure}[htb!]
 \centering
 \includegraphics[scale=0.65,keepaspectratio=true]{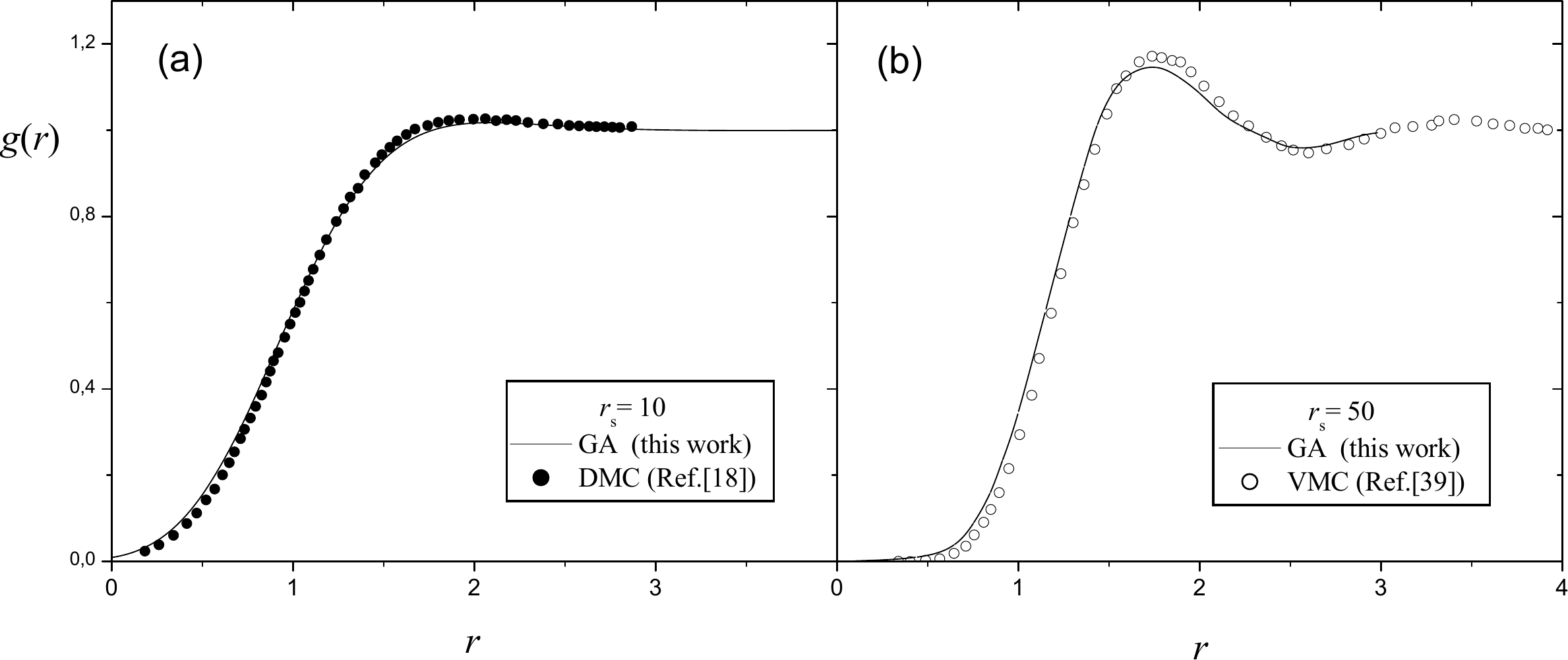}
 \caption{The functions $g(r)$ for the homogenous electron obtained
in this work compared with variational and diffusion Monte Carlo results. (a):
\textit{r}$_{s}$=10; (b): \textit{r}$_{s}$=50.}
 \label{fig:4-5}
\end{figure}

\textbf{Acknowledgments}

Support of this work by Universidad Nacional de La Plata (Project 11/I153),
Universidad Nacional de Rosario (Project 19/VET 47), Consejo Nacional de
Investigaciones Cient\'{\i}ficas y T\'{e}cnicas (PIP 1192) and Agencia
Nacional de Promoci\'{o}n Cient\'{\i}fica y Tecncnol\'{o}gica\ (PICT 00908) of
Argentina is greatly appreciated. F.V. \ and C.M.C. are members of CONICET.

\end{document}